\documentclass[sigconf,natbib=true,anonymous=false]{acmart}
\usepackage{makecell}
\usepackage{multirow}
\AtBeginDocument{%
  \providecommand\BibTeX{{%
    \normalfont B\kern-0.5em{\scshape i\kern-0.25em b}\kern-0.8em\TeX}}}

\begin{document}

\title{Phocus: Picking Valuable Research from a Sea of Citations}

\author{Xinrong Zhang}
\email{zxr19@mails.tsinghua.edu.cn}
\affiliation{%
  \institution{Tsinghua University}
  \city{Beijing}
  \country{China}
  \postcode{100084}
}
\author{Zihou Ren}
\email{rzh20@mails.tsinghua.edu.cn}
\affiliation{%
  \institution{Tsinghua University}
  \city{Beijing}
  \country{China}
  \postcode{100084}
}
\authornote{Authors contributed equally to this research.}
\author{Xi Li}
\email{lixi19@mails.tsinghua.edu.cn}
\affiliation{%
  \institution{Tsinghua University}
  \city{Beijing}
  \country{China}
  \postcode{100084}
}
\authornotemark[1]
\author{Shuqi Liu}
\email{liu-sq19@mails.tsinghua.edu.cn}
\affiliation{%
  \institution{Tsinghua University}
  \city{Beijing}
  \country{China}
  \postcode{100084}
}
\authornotemark[1]
\author{Yunlong Deng}
\email{dengyl20@mails.tsinghua.edu.cn}
\affiliation{%
  \institution{Tsinghua University}
  \city{Beijing}
  \country{China}
  \postcode{100084}
}
\authornotemark[1]
\author{Yadi Xiao}
\email{xyd18@mails.tsinghua.edu.cn}
\affiliation{%
  \institution{Tsinghua University}
  \city{Beijing}
  \country{China}
  \postcode{100084}
}
\authornotemark[1]
\author{Yuxing Han}
\email{yuxinghan@tsinghua-sz.org}
\authornotemark[1]
\affiliation{
 \institution{Tsinghua Shenzhen International Graduate School}
  \city{Shenzhen}
  \country{China}
  \postcode{518057}
}
\author{Jiangtao Wen}
\authornote{Corresponding author}
\email{jtwen@tsinghua.edu.cn}
\affiliation{%
  \institution{Tsinghua University}
  \city{Beijing}
  \country{China}
  \postcode{100084}
}
\renewcommand{\shortauthors}{Zhang and Wen et al.}

\begin{abstract}

The deluge of new papers has significantly blocked the development of academics, which is mainly caused by author-level and publication-level evaluation metrics that only focus on quantity. Those metrics have resulted in several severe problems that trouble scholars focusing on the important research direction for a long time and even promote an impetuous academic atmosphere. To solve those problems, we propose Phocus, a novel academic evaluation mechanism for authors and papers. Phocus analyzes the sentence containing a citation and its contexts to predict the sentiment towards the corresponding reference. Combining others factors, Phocus classifies citations coarsely, ranks all references within a paper, and utilizes the results of the classifier and the ranking model to get the local influential factor of a reference to the citing paper. The global influential factor of the reference to the citing paper is the product of the local influential factor and the total influential factor of the citing paper. Consequently, an author's academic influential factor is the sum of one's contributions to each paper one co-authors.
\end{abstract}





\maketitle

\section{Introduction}
\begin{figure}
\includegraphics[width=8 cm]{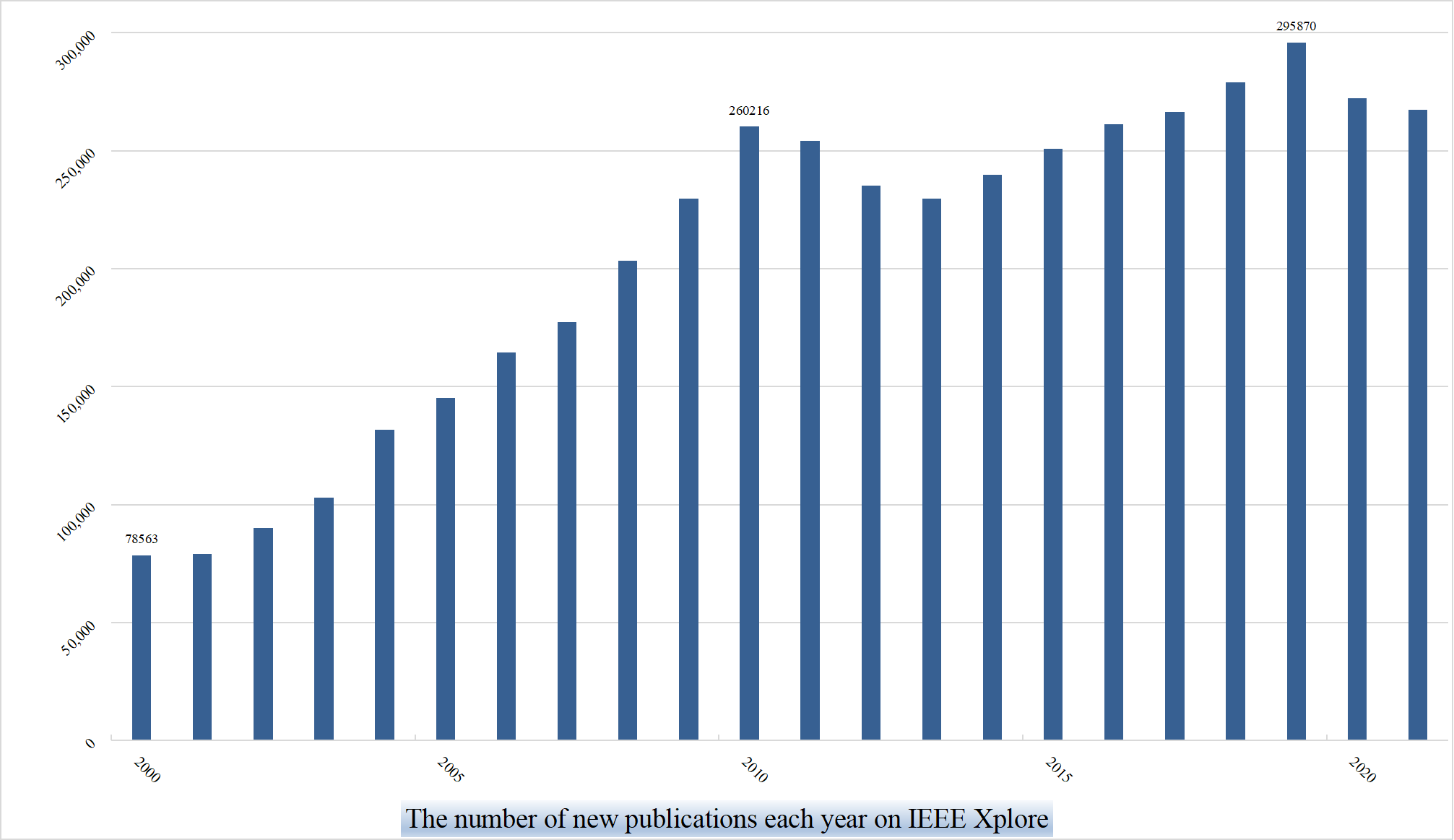}
\caption{the number of new publications on IEEE Xplore each year from 2000 to 2021.}
\label{fig:ieee}
\end{figure}
The number of papers published each year has grown greatly. For example, as shown in Figure~\ref{fig:ieee}, the number of new papers on IEEE Xplore\footnote{\url{https://ieeexplore.ieee.org/Xplore/home.jsp}} increases sharply over the decade.

Paper boom in academic fields results in many severe problems. Cortes et al. \cite{Cortes2021InconsistencyIC} examine 2014 NeurIPS and find that it is not able to pick out excellent researches, and could identify terrible papers. Chu et al. \cite{Chu2021SlowedCP} reveal that too many papers published each year in a field hinder its development. They state this opinion in two aspects. First, researchers are busy coping with a lot of papers, but don't have enough time to fully learn novel ideas; Second, the focused attention on a promising idea might be broken up by the deluge of new ideas.

The reason for the sharp increase in papers is that evaluation metrics for researchers and scholars focus on the number of papers. From the scientific output, research funding, to the evaluation of professional rank, papers play a very important role, and the more papers, the better. However, It is time to make changes. Quantitative metrics could not evaluate the real academic impact of a scholar or a paper. They ignore the essential differences between citations, which is a fatal error. Seglen expresses strong opposition to impact factors that measure the academic influence of journals for committees seldom have the specialist' insights to assess primary researches\cite{Seglen1997WhyTI}.

We propose Phocus, a novel evaluation mechanism for scholars and publications. Phocus analyzes the sentence containing a citation and its contexts to predict the sentiment polarity towards the corresponding reference. Besides, Phocus also considers the total number of citations, the number of citations per sentence, author overlap, and the number of references, similar to \cite{Valenzuela2015IdentifyingMC}. Given those factors above, Phocus uses Naive Bayesian Classifier to divide citations coarsely into 4 categories and utilizes the LambdaMART model to sort all references within a paper. Combining the categories and the ranking results, every reference gets its local influential factor within $ [-1, 1] $, related to the citing paper. The global influential factor of the reference to the citing paper is the product of the local influential factor and the total influential factor of the citing paper. Consequently, an author's academic influential factor is the sum of his contributions to each paper he co-authors.

\section{Related Work}
Our work involves citation classification, aspect-based sentiment analysis, ranking model and evaluation metrics for academics, which will be introduced in subsections below respectively.

\subsection{Citation Classification}

In fact, there are already many kinds of research that have focused on citation classification. For example, Teufel et al. \cite{Teufel2006AutomaticCO} classify citation intents into 12 classes, using simple regular match to extract features. Valenzuela et al. \cite{Valenzuela2015IdentifyingMC} divide citations into 4 classes: highly influential, background, method and results citations, using SVM with an RBF kernel and random forests, taking 13 features into consideration: total number of direct citations, number of direct citations per section, the total number of indirect citations and number of indirect citations per section, author overlap, is considered helpful, citation appears in table and caption, $ 1 / \text{number of references} $, $ \text{number of paper citations} / \text{all citations} $, the similarity between abstracts, PageRank\cite{Page-1999-Pagerank}, number of total citing papers after transitive closure, and field of the cited paper. While Jurgens et al. \cite{Jurgens2016CitationCF} define 7 classes of citation intents: background, motivation, uses, extension, continuation, comparison or contrast, and future, with a Random Forest classifier trained using 4 types of features: structural features, lexical, morphological and grammatical features, field, and usage. Cohan et al. \cite{Cohan2019StructuralSF} propose a multitask model using BiLSTM and attention mechanism to classify citation intents that is the primary task and predict the section where the citation occurs and where a sentence needs a citation that is auxiliary tasks and is used to assist the primary task\footnote{\url{https://github.com/allenai/scicite}}. They categorize intents into 3 classes: background information, method, and result comparison. Besides, Cohan builds a citation intent dataset SciCite. Those works simply classify citations according to intents but ignore the sentiment citing paper towards references, which is vital.

Butt et al. \cite{Butt2015ClassificationOR} utilize Naive-Bayes Classifier to predict the sentiment polarity of a sentence containing a citation and its contexts. Whereas Liu et al. \cite{Liu2017SentimentAO} use averaged word embeddings to represent sentence vectors and to classify sentiment polarities. However, this method generates the overall sentiment of text, rather than the precise sentiment towards the cited paper, which is unable to apply directly.

\subsection{Aspect-based Sentiment Analysis}

Aspect-based sentiment analysis (ABSA) is proposed to define such a task. Usually, ABSA consists of two stages: locating aspects and analyzing sentiment. Some works solve this problem also in a two-stage way, while some jointly.

To detect citation span in Wikipedia, Fetahu et al. \cite{fetahu-etal-2017-fine} propose a sequence classification method using a linear chain CRF to decide which text fragments are covered by a citation at the sub-sentence level. Whereas Kaplan et al. \cite{DainKaplan2016} detect non-explicit citing sentences that surround an explicit citing sentence, utilizing relational, entity, lexical, and grammatical coherence between them. \cite{Ma-2018-AICTS}\cite{Chrysoula-2020-CTSI}even try to find the most relative sentences in reference paper with the citing sentences. Qazvinian and Radev \cite{qazvinian2010identifying} proposed a method based on probabilistic inference to extract non-explicit citing sentences by modelling the sentences in an article and their lexical similarities as a Markov Random Field tuned to detect the patterns that context data create and employ a Belief Propagation mechanism to detect likely context sentences. Abu-Jbara and Radev \cite{abu2012reference} determine the citation block by first segmenting the sentences and then classifying each word in the sentence as being inside or outside the citation block. Finally, they aggregate the labels of all the words contained in a segment to assign a label to the whole segment using three different label aggregation rules(majority label of the words, at least one of the words, or all of them). Kaplan et al. \cite{kaplan2009automatic} proposed a new method based on coreference-chains for extracting citation blocks from research papers.

Given aspects, Sun et al. \cite{Sun2019UtilizingBF} construct an auxiliary sentence from a aspect, and feed the sentence-pair into BERT-based model. Gao et al. \cite{Gao2019TargetDependentSC} utilize three target-dependent variations of the $ BERT_{base} $ model. Bai et al. \cite{Bai-2021-ITSDTSCUGANN} propose a novel relational graph attention network\footnote{\url{https://github.com/muyeby/RGAT-ABSA}}, which integrates typed syntactic dependency information. 

As the errors are cumulated in the pipeline, some researchers explore solutions that detect aspects and classify sentiment jointly. Wang et al. \cite{Wang2010LatentAR} propose a latent aspect rating analysis problem that aims at analyzing reviewers' latent opinions on an entity from several aspects. For a certain entity, they define a set of keywords of aspects and segment reviews into the aspect level. Given aspect segmentation results, they use a novel latent rating regression model to calculate aspect ratings and corresponding weights. However, Wang et al. ignore the inter-dependencies between words and sentences, which causes great information loss. This class problem is also called aspect-based sentiment analysis (ABSA). Ruder et al. \cite{Ruder2016AHM} proposes a hierarchical bidirectional LSTM to model the inter-dependencies of sentences within a review. The aspect is represented by the average of its entity and attribute embeddings. Hoang et al. \cite{Hoang2019AspectBasedSA} propose to use a sentence pair classifier model from BERT\cite{Devlin2019BERTPO} to solve ABSA at sentence and text levels. Hu et al. \cite{hu-etal-2019-open} propose a span-based extract-then-classify framework based on BERT\footnote{\url{https://github.com/huminghao16/SpanABSA}}. Xu et al. \cite{Xu2019BERTPF} build a dataset, ReviewRC\footnote{\url{https://howardhsu.github.io/dataset/}}, and extend BERT with an extra tasking-specific layer to tune each task. Wallaart et al. \cite{Wallaart2019AHA} propose a two-stage algorithm to solve the ABSA for restaurant reviews: predicting the sentiment with a lexicalized domain ontology, and using a neural network with a rotatory attention mechanism (LCR-Rot) as a backup algorithm. The order of rotatory attention mechanism operation is changed and the rotatory attention mechanism is iterated multiple times. Trusca et al. extend \cite{Wallaart2019AHA} with deep contextual word embeddings and add an extra attention layer to its high-level representations\cite{Trusca2020AHA}. To address the imbalance issue and utilize the interaction between aspect terms, Luo et al. \cite{luo-etal-2020-grace} propose a gradient harmonized and cascaded labelling model based on BERT. Chen et al. \cite{chen-etal-2020-joint-aspect} utilize directional graph convolutional networks to perform end-to-end ABSA task.




\subsection{Ranking Model}

The ranking model is based on LambdaMART, which is the boosted tree version of LambdaRank\cite{Burges2006LearningTR}. This algorithm solves the gradients of non-smooth cost functions used in ranking models. Burges et al. \cite{Burges2010FromRT} give a review on RankNet, LambdaRank, and LambdaMART. 

To illustrate the ranking network, we use $ c_{ij} $ to denote the $ j $-th citation of the $ i $-th reference paper. Our ranking network receives an matrix of shape $ (\sum_i n\_cit_i, 4) $, where 4 stands for the feature quaternion of (au\_overlap, n\_cit, cit\_word, sen\_label). Among which cit\_word is calculated as the total number of words in $ context\_a + sentence + context\_b $. The network calculate a score $ s_{ij} $ on each time of citation $ c_{ij} $ individually, averaging on duplicate citations to get the score of each reference paper $ s_i = \frac{1}{n\_cit_i} \sum_j s_{ij} $. Then $ s_i $ is used to rank all the reference paper, outputting $ r_i $.


\subsection{Evaluation Metrics}

In the academic field, there are journal-level, author-level and paper-level metrics that measure their impacts. 

The Impact Factor (IF)\cite{Milstead1980CitationIT} and CiteScore\footnote{\url{https://service.elsevier.com/app/answers/detail/a_id/14880/supporthub/scopus/}} are used to measure the impact of a journal based on the number of times articles cited during a fixed period published by the journal. Besides, Journal Citation Reports (JCR) give ranking for journals\footnote{\url{https://jcr.clarivate.com/jcr/home}}, Eigenfactor scores\cite{Bergstrom2007EigenfactorMT} measure how likely a journal is to be used, and SCImago Journal Rank (SJR)\cite{gonzalezpereira2009sjr} regards the citations issued by more import journals as more important than those issued by less important ones. Whereas Source Normalized Impact per Paper (SNIP)\cite{MOED2010265} indicates that a single citation is much more important in subject areas where citations are less, and vice versa.

Author-level metrics include h-index, g-index, i10-index and so on. H-index also called index $ h $,  is proposed by Jorge E. Hirsch\cite{Hirsch2005H-index}, and its definition is the number of papers with citation numbers higher or equal to $ h $. The g-index is defined as the largest number such that the top $ g $ articles received together at least $ g^{2} $ citations\cite{Egghe2006G-index}. Google Scholar proposes the i10-index that is the number of a publication with at least 10 citations. Those metrics are derived from citations and do not reveal the truth among citations.

Paper-level metrics are usually the number of citations. Especially, Semantic Scholar makes the first step towards citation classification. It divided citations into 4 classes: highly influential, background, method and results citations\cite{Valenzuela2015IdentifyingMC}, using SVM with an RBF kernel and random forests. The features Semantic Scholar use are the total number of direct citations, number of direct citations per section, the total number of indirect citations and number of indirect citations per section, author overlap, is considered helpful, citation appears in table and caption, $ 1 / \text{number of references} $, $ \text{number of paper citations} / \text{all citations} $, the similarity between abstracts, PageRank\cite{Page-1999-Pagerank}, number of total citing papers after transitive closure, and field of the cited paper.

\section{Methodology}
\begin{figure}
\label{fig:overview}
\includegraphics[width=8.5 cm]{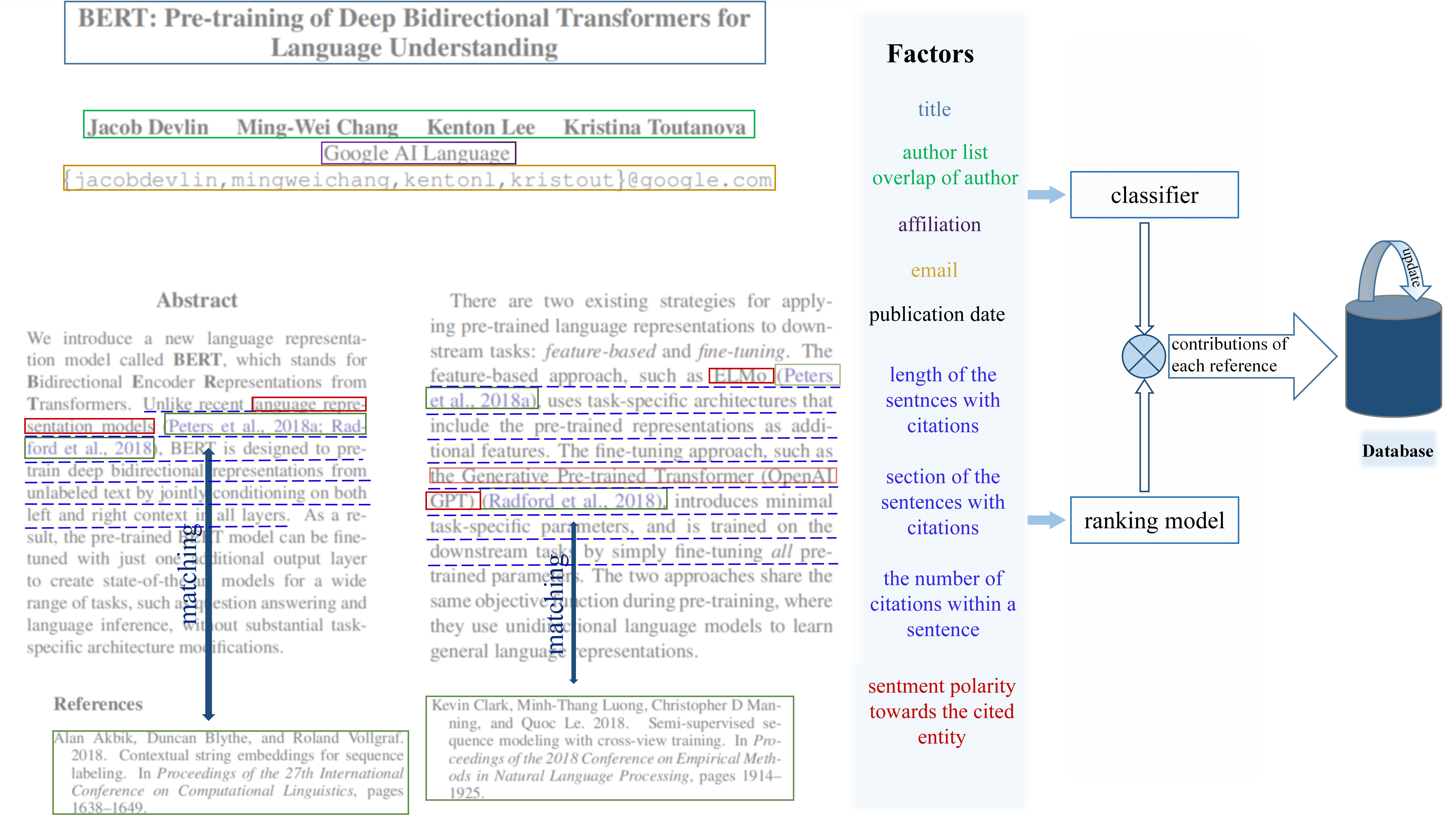}
\caption{the overview of Phocus.}
\end{figure}
As shown in Figure~\ref{fig:overview}, our algorithm consists of 4 stages: pre-processing, calculating factors, evaluating contribution, and propagating influential factors. In pre-processing stage, we clean raw data, and obtain simple factors. Complex factors, like sentiment polarity are calculated in second stage. When get all factors needed, we classify citations into four classes and rank all references, and figure out the local contribution factor of each reference. We initialize all new paper to the database with an academic influential factor 1.0, and propagate its impact on references iteratively. The factors extracted from papers are listed out in Table~\ref{tab:allfactors}
\subsection{Pre-processing}
\begin{table}
  \caption{factor list}
  \label{tab:allfactors}
  \begin{tabular}{p{1.5 cm}|p{3.8 cm}|p{2 cm}}
    \toprule
    Name&Definition&Ranges\\
    \midrule
    cit\_id & \makecell[l]{reference number of a\\paper in the reference list} & positive integer \\
    cit\_title & \makecell[l]{title of a reference}& string\\
    cit\_author & \makecell[l]{authors of cit\_title} & list of authors\\
    cit\_year & \makecell[l]{publish year of cit\_title} & year\\
    \midrule
    au\_overlap& \makecell[l]{overlap between authors\\of cit\_title and citing paper} & [0, 1]\\
    sent\_id & \makecell[l]{id of a sentence} & natural number\\
    sec\_id& \makecell[l]{section id of a sentence}&\makecell[l]{0: related work\\ introduction\\ 1: main body\\ 2: conclusion} \\
    n\_cit& \makecell[l]{time of cit\_id cited\\ in citing paper}& natural number\\
    \midrule
    cit\_text&\makecell[l]{text of the sentence that\\ contains the cit\_id} & string\\
    context\_a & \makecell[l]{related sentences previous\\ to cit\_text} & string\\
    context\_b& \makecell[l]{related sentences behind\\ to cit\_text} &string\\
    sen\_label& \makecell[l]{the sentiment citing paper\\towards cit\_ id} & \makecell[l]{-1: negative\\ 0: neutral\\ 1: positive}\\
  \bottomrule
\end{tabular}
\end{table}

Given a paper of string format, a series of steps process the raw data for the next stage: parsing, segmentation, and matching. Paring is aimed at dividing the input text into title, authors, sections, and references. We utilize flari\footnote{\url{https://pypi.org/project/flair/}} to parse the title, authors and publish year of the input paper and its references. We segment the input paper into two-level: section level and sentence level. Section segmentation is based on keywords matching and classified into three categories: 0 representing related work, introduction or other background citation; 1 representing main body including methodology, experiments and so on; 2 representing conclusion and other parts. Sentences are segmented using regular expression matching and are then labelled by their ID according to their appearing order. Reference parsing generates title, authors, publish year and even their citation markers in the paper. Given that information, we locate citations in each sentence and match citation markers with their corresponding reference papers. Then we could easily get the factor n\_cit and cit\_text. Factor au\_overlap is calculated according to the following equation:
\begin{equation}
\textstyle au\_overlap = 2 \times \frac{A \cap B }{|A| + |B|}
\end{equation}
where A is the author set of citing paper, and B is the author set of reference paper. 

\subsection{Calculating Factors}
There are still three factors unsolved: context\_a, context\_b, and sen\_label. We obtain context\_a, context\_b with BERT, and propose a novel aspect-based sentiment analysis algorithm to classify citation sentiment.

We fine-tune BERT on a manually annotated dataset containing over 1,000 sentence pairs labelled as "related" or "irrelevant". Each sentence pair is generated from a single academic paper. We get an accuracy of 94.5\% on the evaluation dataset. To obtain the context of cit\_context, we apply the above classifier iteratively on sentence pair $(S[sent\_id - i], S[sent\_id])$ ($S$ representing the list of all sentences in the paper) where $i$ increases from 1. Once an "irrelevant" pair is reported, the iteration is aborted and we take $S[sent\_id - i:sent\_id]$ as context\_a. Another stopping criterion is that $S[sent\_id - i]$ should always be in the same paragraph with $S[sent\_id]$. A similar procedure is performed on $(S[sent\_id + i], S[sent\_id])$ to get context\_b.

\subsection{Evaluating Contribution}
After gathering all needed factors, we train a classifier to categorize citation into 4 classes: very important, important, neutral, and terrible. And we also train a ranking model to predict the related order of references in terms of their contributions to the paper.
\begin{table}
\caption{the classifying standards of Phocus.}
\begin{tabular}{p{1 cm}p{7 cm}}
\toprule
Label&Description \\
\midrule
3 & \makecell[l]{extending the work; highly influenced by the work}\\
2 & using the work \\
1 & related work \\
0 & negative sentiment towards the work\\
\bottomrule
\end{tabular}
\label{tab:class-standard}
\end{table}
First, we classify citations into four categories with a Naive Bayesian classifier. The classifying standards are shown in Table~\ref{tab:class-standard}, and a larger number of labels represents more contributions.

The ranking model is based on LambdaMART, which is the boosted tree version of LambdaRank\cite{Burges2006LearningTR}. This algorithm solves the gradients of non-smooth cost functions used in ranking models. Burges et al. \cite{Burges2010FromRT} give a review on RankNet, LambdaRank, and LambdaMART. 

Based on the classes and order of references, we project them into $ [0, 1] $ to get their influential factors.

\subsection{Propagating Influential Factors}
\begin{figure}
\includegraphics[width=6 cm]{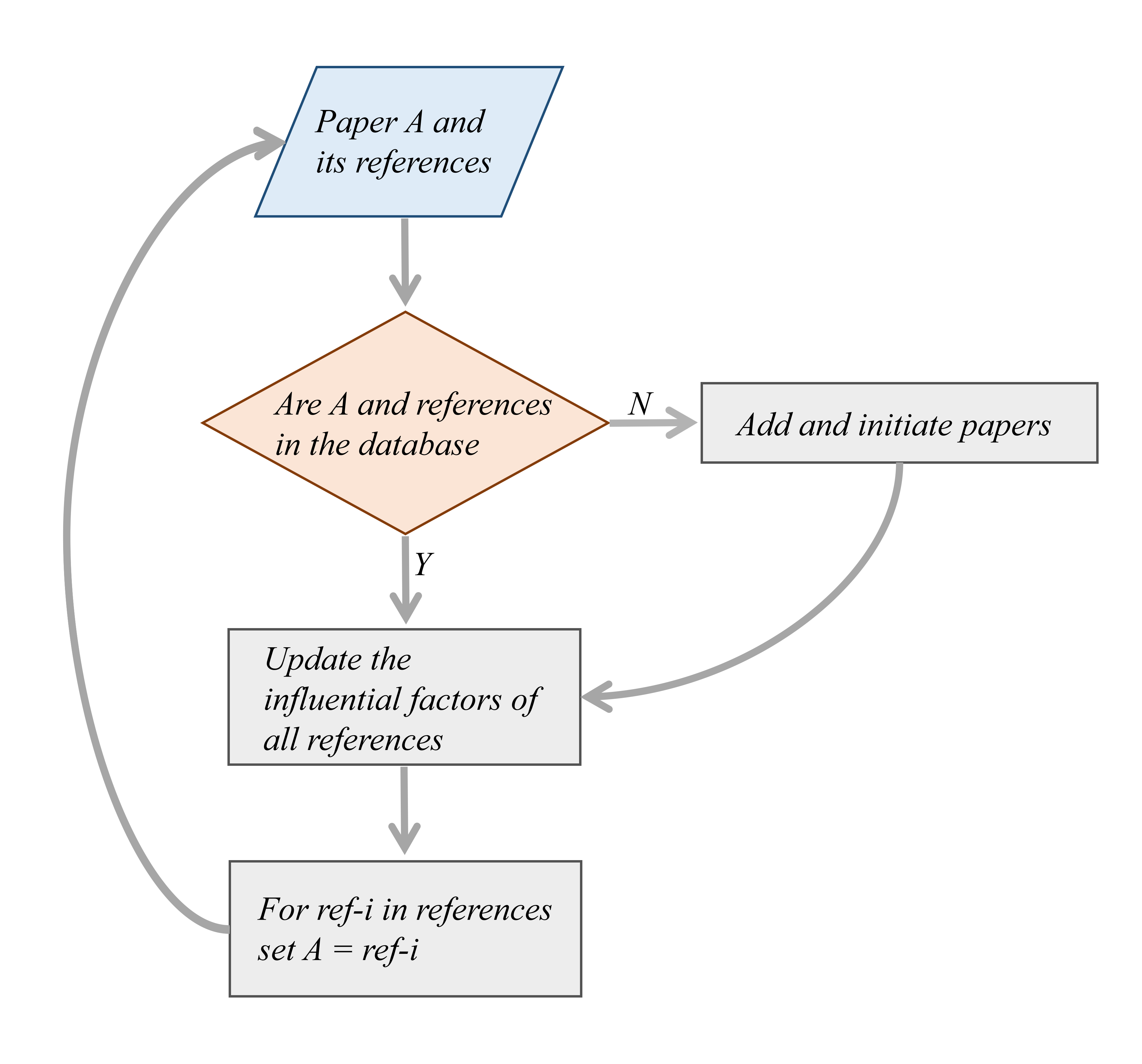}
\caption{the propagation rules of influential factors}
\label{fig:factors-prop}
\end{figure}
Given a list of references and their influential factors of the citing paper, we design some rules to propagate their influence. The main idea is shown in Figure~\ref{fig:factors-prop}.

$ A $ denote a citing paper with academic influential factor $ AF_{A} $ initialized as 1, set $ R_{A} $, $ IF_{A}^{l}$ denote all references of $ A $, and their corresponding local contribution to $ A $, and $ IF_{Ai}^{l} \in [-1, 1] $ is the local contribution of reference i to $A$. $ C_{A} $ is the set of all papers that cite $ A $, and for $ j \in C_{A} $,  $ IF_{jA}^{l} \in [-1, 1] $ is A's local contribution to $ j $. Then, the academic influential factor of $ A $ is:
\begin{equation}
AF_{A} = \sum_{\substack{j \in C_{A}}} AF_{j}IF_{jA}^{l}
\end{equation}
For author $ a $ who publishes a set of papers $ P_{a} $, and his contribution to paper $ i \in P_{a} $ is $ C_{ia} \in [0, 1] $, his academic influential factor is:
\begin{equation}
AF_{a} = \sum_{\substack{i \in P_{a}}} C_{ia}AF_{i}
\end{equation}
For paper $ A $, and its $ N $ authors, $ \sum_{\substack{i}}^{\substack{N}} C_{Ai} \equiv 1 $.
There are two problems to prove to ensure that our method is logical. The first one is margin effects. And the second one the propagation rules.

\section{Experiments}
We conduct several experiments to demonstrate our new metrics that measure the influential factors of an individual scientist or scholar and the citation impact of the publications.

As the influential factor of a paper is the weighted sum of all papers that cite it and its corresponding contribution to them, the final and full network of paper and network should be constructed. However, we cannot complete this job yet out of no access to some databases, not enough time or computational resources. We will select some scholars and their publications as targets, and utilize primary citation and secondary citation relationships. Besides, we also compare our modules to other state-of-art algorithms to show the improvement we achieve.

\subsection{Peer Comparison}
Scholar and their publications. Let Scholar Y denote some scholar. We will show the difference between Scholar Y and the Turing Award winner Pat. Hanrahan\footnote{\url{https://scholar.google.com/citations?hl=zh-CN&user=RzEnQmgAAAAJ}}. As we emphasize, Pat. Hanrahan is much more influential than scholar Y is not only for that he wins Turing Award, but also is based on solid statistics of citations. For example, He et al. \cite{he2015deep} take one paper of scholar Y as a baseline that performs only better than one baseline among eleven. Table~\ref{tab:impact-comparision} shows evaluation results of scholar Y and Pat. Hanrahan on Aminer\footnote{\url{https://www.aminer.cn/}}, Google Scholar\footnote{\url{https://scholar.google.com/}}, Semantic Scholar\footnote{\url{https://www.semanticscholar.org/}} and Phocus.
\begin{table}
\small
\caption{statistics of Y and Hanrahan}
\label{tab:statistics}
\begin{tabular}{p{1cm}| p{1.1cm}p{0.8cm}|p{0.8cm}|p{1.1cm}p{0.8cm}}
\toprule
\multicolumn{1}{c|}{Scholar} & \multicolumn{2}{c|}{Aminer}&\multicolumn{1}{c|}{Google Scholar}&\multicolumn{2}{c}{Semantic Scholar} \\
 &publications&citations&citations&publications&citations\\
\midrule
Y & 1146 & 77903 & 78663& 771 & 59679  \\
\midrule
Hanrahan &381&52214&50568&315&56383\\
\bottomrule
\end{tabular}
\end{table}
Table~\ref{tab:statistics} lists the number of publications and citations of scholar Y and Pat. Hanrahan. It's obviously that scholar Y is more productive than Pat. Hanrahan.  However, those numbers covers up some significant truths that not all papers are equal influential and not all citations mean agreement with the cited ones. 
\begin{table}
\footnotesize
\caption{evaluation results from several platforms}
\label{tab:impact-comparision}
\begin{tabular}{p{0.8cm}p{0.4cm}p{0.4cm}p{0.4cm}p{0.4cm}p{0.4cm}p{0.4cm}p{0.4cm}}
\toprule
\multicolumn{1}{c}{Scholar} & \multicolumn{2}{c}{Aminer}&\multicolumn{2}{c}{Google Scholar}&\multicolumn{2}{c}{Semantic Scholar} & \multicolumn{1}{c}{\makecell[l]{Phocus \\(Primary)}} \\
 &h&g&h&i10&h&HIC&\\
\midrule
Y &  \textbf{131} & \textbf{258} & \textbf{123} & \textbf{723} &\textbf{119} &\textbf{5843}&0.40\\
\midrule
Hanrahan &97&228&93&200&88&3741&\textbf{0.52} \\
\bottomrule
\end{tabular}
\end{table}
where h represents h-index, g represents g-index, i10 means i10-index, and HIC is the number of highly influential citations. H-index, also called index $ h$,  is proposed by Jorge E. Hirsch\cite{Hirsch2005H-index}, and its definition is the number of papers with citation number higher or equal to $ h $ . The g-index is defined as the largest number such that the top $ g $ articles received together at least $ g^{2} $ citations\cite{Egghe2006G-index}. Google Scholar proposes i10-index that is the number of a publication with at least 10 citations. Those metrics are derived from citations and do not reveal the truth among citations. Semantic Scholar makes the first step towards citation classification. It divided citations into 4 classes: highly influential, background, method and results citations\cite{Valenzuela2015IdentifyingMC}, using SVM with a RBF kernel and random forests. The features Semantic Scholar use are total number of direct citations, number of direct citations per section, total number of indirect citations and number of indirect citations per section, author overlap, is considered helpful, citation appears in table and caption, $ 1 / \text{number of references} $, $ \text{number  of  paper  citations} / \text{all  citations} $, similarity between abstracts, PageRank\cite{Page-1999-Pagerank}, number of total citing papers after transitive closure, and field of the cited paper.
\begin{figure}
\includegraphics[width=6cm]{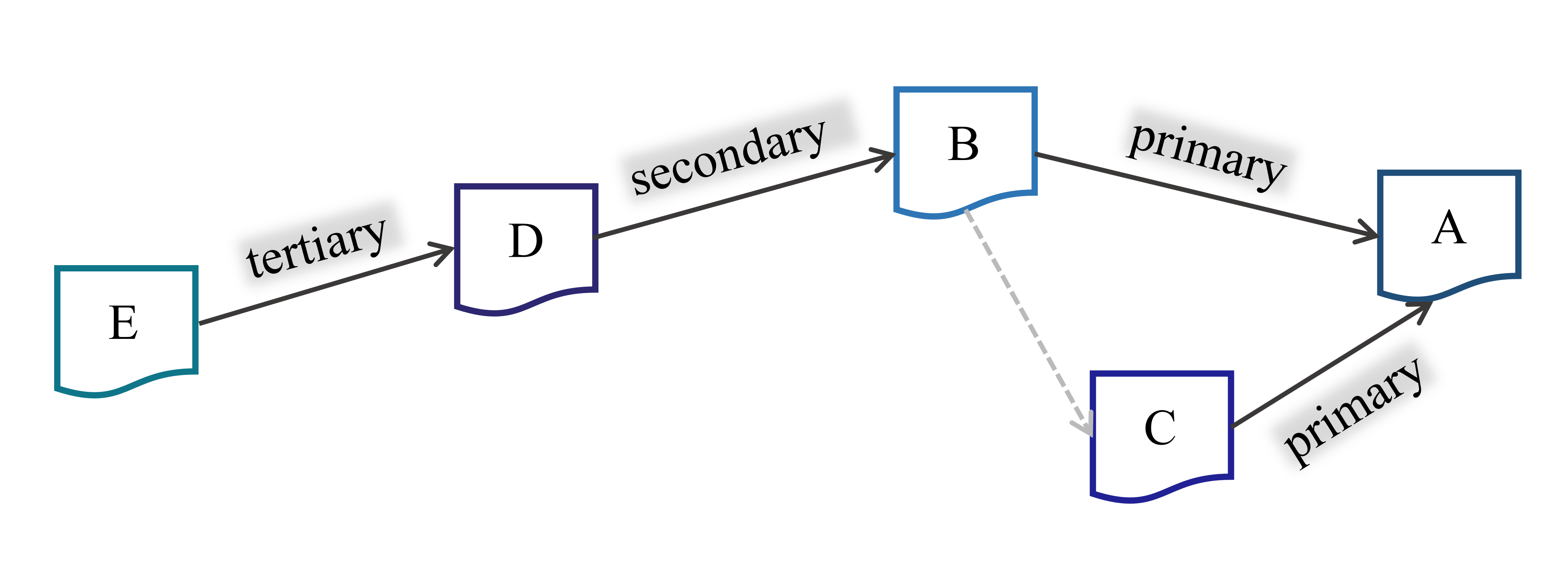}
\caption{for paper A, paper B and C cite it directly, called primary citations, D to A is secondary and E to A is tertiary.}
\label{fig:citationrelation}
\end{figure}
We collect XX papers that cite scholar Y from 78663, and XX papers that cite Patrick Hanrahan from 56383. Only utilizing primary citations, we get the global academic influential factors of scholar Y and Patrick Hanrahan is 0.40 and 0.52 respectively. Figure~\ref{fig:citationrelation}

\subsection{Mathematical Invariance}
To verify the model, we conduct a series of experiments to prove it's reasonable.

First, given a set of references within a paper, removing anyone reference from the set won't change the related order of left references. And when removing a reference at a time, the left references also keep related orders.

Also, the final score should be stable and insensitive to propagating order under a certain paper pool. Our strategy starts from a default influential factor 1.0, traversing through each paper and updating the influential factor successively. It is proven through experiments that regardless of the updating order, the final score of each paper remains the same.

\subsection{Citation Span}
We conduct some experiments guided by \cite{abu2012reference} as our baseline. We annotate the citation span for about 345 citing sentences as our data set to train and test the baseline model.

First, we use the tokenizer tool that SpaCy\footnote{\url{https://spacy.io/}} provides to segment the text of each citing sentence into tokens, and use tagger and parser tool to assign part-of-speech-tags and dependency labels to each token.

Then, we extract features listed in Table~\ref{citation_span_features}. as the input of the baseline model. The training is performed using SVM, Logistic Regression, and CRF, respectively. We use 10-fold cross-validation for training and testing.

\begin{table}
\caption{features used for citation span}
\label{citation_span_features}
\begin{tabular}{@{}l|l@{}}
\toprule
\textbf{Feature} & \textbf{Description} \\ \midrule
distance & \begin{tabular}[c]{@{}l@{}}The distance (in words) between the word \\ and the target citaion.\end{tabular} \\ \midrule
position & \begin{tabular}[c]{@{}l@{}}This feature takes the value 1 if the word \\ comes before the target citation, and 0 otherwise.\end{tabular} \\ \midrule
segment & \begin{tabular}[c]{@{}l@{}}After splitting the sentence into segments \\ by punctuation and coordination conjunctions, \\ this feature takes the value 1 if the word occurs \\ in the same segment with the target reference, \\ and 0 otherwise.\end{tabular} \\ \midrule
pos\_tag & \begin{tabular}[c]{@{}l@{}}The part of speech tag of the word, the word \\ before, and the word after.\end{tabular} \\ \midrule
dTreeDistance & \begin{tabular}[c]{@{}l@{}}Length of the shortest dependency path (in \\ the dependency parse tree) that connects the \\ word to the target reference or its representative.\end{tabular} \\ \midrule
lca & \begin{tabular}[c]{@{}l@{}}The type of the node in the dependency parse \\ tree that is the least common ancestor of the \\ word and the target reference.\end{tabular} \\ \bottomrule
\end{tabular}
\end{table}

Table~\ref{citation_span_results} lists the precision, recall, and F1 for the three model.
\begin{table}
\caption{results for three different models for citation span}
\label{citation_span_results}
\begin{tabular}{@{}llll@{}}
\toprule
\textbf{Model} & \textbf{Precision} & \textbf{Recall} & \textbf{F1} \\ \midrule
SVM & 0.78 & 0.56 & 0.65 \\
LR & 0.68 & 0.67 & 0.67 \\
CRF & 0.65 & 0.64 & 0.64 \\ \bottomrule
\end{tabular}
\end{table}


\section{Results}
As shown in Table~\ref{tab:impact-comparision}, Phocus figures out that the global academic influential factors of scholar Y and Patrick Hanrahan are 0.40 and 0.52 respectively, and Patrick Hanrahan is 30\% higher than scholar Y. It's the results that only utilize primary citation data. While the evaluation results from Aminer, Google Scholar and even Semantic Scholar shows that scholar Y is more productive and influential than Patrick Hanrahan.

\section{Conclusion}

In this paper, we come up with Phocus, a novel set of academic evaluation metrics for authors and publications based on citation judgements that utilize aspect-based sentiment analysis. To verify our evaluation mechanism, peer comparison and ablation studies have been conducted. The results show that our metrics are able to identify the truly worthiness of a paper or a scholar, which is difficult to citation times based metrics, like h-index, g-index and others. 

Phocus still need improvements. As shown in Section Experiments, we only use primary citation data, which is not enough to fully prove the reliability of Phocus. Besides, using more data such as secondary and tertiary citations could further reflect the gaps between scholars and between metrics. There are still many problems unsolved, such as “citation circles” (groups of researchers who cite one another’s work), and self-citation.

\bibliographystyle{ACM-Reference-Format}
\bibliography{ref}


\begin{thebibliography}{39}


\ifx \showCODEN    \undefined \def \showCODEN     #1{\unskip}     \fi
\ifx \showDOI      \undefined \def \showDOI       #1{#1}\fi
\ifx \showISBNx    \undefined \def \showISBNx     #1{\unskip}     \fi
\ifx \showISBNxiii \undefined \def \showISBNxiii  #1{\unskip}     \fi
\ifx \showISSN     \undefined \def \showISSN      #1{\unskip}     \fi
\ifx \showLCCN     \undefined \def \showLCCN      #1{\unskip}     \fi
\ifx \shownote     \undefined \def \shownote      #1{#1}          \fi
\ifx \showarticletitle \undefined \def \showarticletitle #1{#1}   \fi
\ifx \showURL      \undefined \def \showURL       {\relax}        \fi
\providecommand\bibfield[2]{#2}
\providecommand\bibinfo[2]{#2}
\providecommand\natexlab[1]{#1}
\providecommand\showeprint[2][]{arXiv:#2}

\bibitem[\protect\citeauthoryear{Abu-Jbara and Radev}{Abu-Jbara and
  Radev}{2012}]%
        {abu2012reference}
\bibfield{author}{\bibinfo{person}{Amjad Abu-Jbara} {and}
  \bibinfo{person}{Dragomir Radev}.} \bibinfo{year}{2012}\natexlab{}.
\newblock \showarticletitle{Reference scope identification in citing
  sentences}. In \bibinfo{booktitle}{\emph{Proceedings of the 2012 Conference
  of the North American Chapter of the Association for Computational
  Linguistics: Human Language Technologies}}. \bibinfo{pages}{80--90}.
\newblock


\bibitem[\protect\citeauthoryear{Bai, Liu, and Zhang}{Bai
  et~al\mbox{.}}{2021}]%
        {Bai-2021-ITSDTSCUGANN}
\bibfield{author}{\bibinfo{person}{Xuefeng Bai}, \bibinfo{person}{Pengbo Liu},
  {and} \bibinfo{person}{Yue Zhang}.} \bibinfo{year}{2021}\natexlab{}.
\newblock \showarticletitle{Investigating Typed Syntactic Dependencies for
  Targeted Sentiment Classification Using Graph Attention Neural Network}.
\newblock \bibinfo{journal}{\emph{IEEE/ACM Trans. Audio, Speech and Lang.
  Proc.}}  \bibinfo{volume}{29} (\bibinfo{date}{jan} \bibinfo{year}{2021}),
  \bibinfo{pages}{503–514}.
\newblock
\showISSN{2329-9290}
\urldef\tempurl%
\url{https://doi.org/10.1109/TASLP.2020.3042009}
\showDOI{\tempurl}


\bibitem[\protect\citeauthoryear{Bergstrom}{Bergstrom}{2007}]%
        {Bergstrom2007EigenfactorMT}
\bibfield{author}{\bibinfo{person}{Carl~T. Bergstrom}.}
  \bibinfo{year}{2007}\natexlab{}.
\newblock \showarticletitle{Eigenfactor Measuring the value and prestige of
  scholarly journals}.
\newblock \bibinfo{journal}{\emph{College \& Research Libraries News}}
  \bibinfo{volume}{68} (\bibinfo{year}{2007}), \bibinfo{pages}{314--316}.
\newblock


\bibitem[\protect\citeauthoryear{Burges}{Burges}{2010}]%
        {Burges2010FromRT}
\bibfield{author}{\bibinfo{person}{Christopher J.~C. Burges}.}
  \bibinfo{year}{2010}\natexlab{}.
\newblock \showarticletitle{From RankNet to LambdaRank to LambdaMART: An
  Overview}.
\newblock


\bibitem[\protect\citeauthoryear{Burges, Ragno, and Le}{Burges
  et~al\mbox{.}}{2006}]%
        {Burges2006LearningTR}
\bibfield{author}{\bibinfo{person}{Christopher J.~C. Burges},
  \bibinfo{person}{Robert~J. Ragno}, {and} \bibinfo{person}{Quoc~V. Le}.}
  \bibinfo{year}{2006}\natexlab{}.
\newblock \showarticletitle{Learning to Rank with Nonsmooth Cost Functions}. In
  \bibinfo{booktitle}{\emph{NIPS}}.
\newblock


\bibitem[\protect\citeauthoryear{Butt, Rafi, Jamal, Rehman, Alam, and
  Alam}{Butt et~al\mbox{.}}{2015}]%
        {Butt2015ClassificationOR}
\bibfield{author}{\bibinfo{person}{Bilal~Hayat Butt}, \bibinfo{person}{Muhammad
  Rafi}, \bibinfo{person}{Arsal Jamal}, \bibinfo{person}{Raja Sami~Ur Rehman},
  \bibinfo{person}{Syed Muhammad~Zubair Alam}, {and}
  \bibinfo{person}{Muhammad~Bilal Alam}.} \bibinfo{year}{2015}\natexlab{}.
\newblock \showarticletitle{Classification of Research Citations (CRC)}. In
  \bibinfo{booktitle}{\emph{CLBib@ISSI}}.
\newblock


\bibitem[\protect\citeauthoryear{Chen, Tian, and Song}{Chen
  et~al\mbox{.}}{2020}]%
        {chen-etal-2020-joint-aspect}
\bibfield{author}{\bibinfo{person}{Guimin Chen}, \bibinfo{person}{Yuanhe Tian},
  {and} \bibinfo{person}{Yan Song}.} \bibinfo{year}{2020}\natexlab{}.
\newblock \showarticletitle{Joint Aspect Extraction and Sentiment Analysis with
  Directional Graph Convolutional Networks}. In
  \bibinfo{booktitle}{\emph{Proceedings of the 28th International Conference on
  Computational Linguistics}}. \bibinfo{publisher}{International Committee on
  Computational Linguistics}, \bibinfo{address}{Barcelona, Spain (Online)},
  \bibinfo{pages}{272--279}.
\newblock
\urldef\tempurl%
\url{https://doi.org/10.18653/v1/2020.coling-main.24}
\showDOI{\tempurl}


\bibitem[\protect\citeauthoryear{Chu and Evans}{Chu and Evans}{2021}]%
        {Chu2021SlowedCP}
\bibfield{author}{\bibinfo{person}{Johan S.~G. Chu} {and}
  \bibinfo{person}{James~A. Evans}.} \bibinfo{year}{2021}\natexlab{}.
\newblock \showarticletitle{Slowed canonical progress in large fields of
  science}.
\newblock \bibinfo{journal}{\emph{Proceedings of the National Academy of
  Sciences of the United States of America}}  \bibinfo{volume}{118}
  (\bibinfo{year}{2021}).
\newblock


\bibitem[\protect\citeauthoryear{Cohan, Ammar, van Zuylen, and Cady}{Cohan
  et~al\mbox{.}}{2019}]%
        {Cohan2019StructuralSF}
\bibfield{author}{\bibinfo{person}{Arman Cohan}, \bibinfo{person}{Waleed
  Ammar}, \bibinfo{person}{Madeleine van Zuylen}, {and} \bibinfo{person}{Field
  Cady}.} \bibinfo{year}{2019}\natexlab{}.
\newblock \showarticletitle{Structural Scaffolds for Citation Intent
  Classification in Scientific Publications}.
\newblock \bibinfo{journal}{\emph{ArXiv}}  \bibinfo{volume}{abs/1904.01608}
  (\bibinfo{year}{2019}).
\newblock


\bibitem[\protect\citeauthoryear{Cortes and Lawrence}{Cortes and
  Lawrence}{2021}]%
        {Cortes2021InconsistencyIC}
\bibfield{author}{\bibinfo{person}{Corinna Cortes} {and} \bibinfo{person}{Neil
  Lawrence}.} \bibinfo{year}{2021}\natexlab{}.
\newblock \showarticletitle{Inconsistency in Conference Peer Review: Revisiting
  the 2014 NeurIPS Experiment}.
\newblock \bibinfo{journal}{\emph{ArXiv}}  \bibinfo{volume}{abs/2109.09774}
  (\bibinfo{year}{2021}).
\newblock


\bibitem[\protect\citeauthoryear{Devlin, Chang, Lee, and Toutanova}{Devlin
  et~al\mbox{.}}{2019}]%
        {Devlin2019BERTPO}
\bibfield{author}{\bibinfo{person}{Jacob Devlin}, \bibinfo{person}{Ming-Wei
  Chang}, \bibinfo{person}{Kenton Lee}, {and} \bibinfo{person}{Kristina
  Toutanova}.} \bibinfo{year}{2019}\natexlab{}.
\newblock \showarticletitle{BERT: Pre-training of Deep Bidirectional
  Transformers for Language Understanding}.
\newblock \bibinfo{journal}{\emph{ArXiv}}  \bibinfo{volume}{abs/1810.04805}
  (\bibinfo{year}{2019}).
\newblock


\bibitem[\protect\citeauthoryear{Egghe}{Egghe}{2006}]%
        {Egghe2006G-index}
\bibfield{author}{\bibinfo{person}{Leo Egghe}.}
  \bibinfo{year}{2006}\natexlab{}.
\newblock \showarticletitle{Theory and practise of the g-index}.
\newblock \bibinfo{journal}{\emph{Scientometrics}}  \bibinfo{volume}{69}
  (\bibinfo{year}{2006}), \bibinfo{pages}{131--152}.
\newblock


\bibitem[\protect\citeauthoryear{Fetahu, Markert, and Anand}{Fetahu
  et~al\mbox{.}}{2017}]%
        {fetahu-etal-2017-fine}
\bibfield{author}{\bibinfo{person}{Besnik Fetahu}, \bibinfo{person}{Katja
  Markert}, {and} \bibinfo{person}{Avishek Anand}.}
  \bibinfo{year}{2017}\natexlab{}.
\newblock \showarticletitle{Fine Grained Citation Span for References in
  {W}ikipedia}. In \bibinfo{booktitle}{\emph{Proceedings of the 2017 Conference
  on Empirical Methods in Natural Language Processing}}.
  \bibinfo{publisher}{Association for Computational Linguistics},
  \bibinfo{address}{Copenhagen, Denmark}, \bibinfo{pages}{1990--1999}.
\newblock
\urldef\tempurl%
\url{https://doi.org/10.18653/v1/D17-1212}
\showDOI{\tempurl}


\bibitem[\protect\citeauthoryear{Gao, Feng, Song, and Wu}{Gao
  et~al\mbox{.}}{2019}]%
        {Gao2019TargetDependentSC}
\bibfield{author}{\bibinfo{person}{Zhengjie Gao}, \bibinfo{person}{Ao Feng},
  \bibinfo{person}{Xinyu Song}, {and} \bibinfo{person}{Xi Wu}.}
  \bibinfo{year}{2019}\natexlab{}.
\newblock \showarticletitle{Target-Dependent Sentiment Classification With
  BERT}.
\newblock \bibinfo{journal}{\emph{IEEE Access}}  \bibinfo{volume}{7}
  (\bibinfo{year}{2019}), \bibinfo{pages}{154290--154299}.
\newblock


\bibitem[\protect\citeauthoryear{Gonzalez-Pereira, Guerrero-Bote, and
  Moya-Anegon}{Gonzalez-Pereira et~al\mbox{.}}{2009}]%
        {gonzalezpereira2009sjr}
\bibfield{author}{\bibinfo{person}{Borja Gonzalez-Pereira},
  \bibinfo{person}{Vicente Guerrero-Bote}, {and} \bibinfo{person}{Felix
  Moya-Anegon}.} \bibinfo{year}{2009}\natexlab{}.
\newblock \bibinfo{title}{The SJR indicator: A new indicator of journals'
  scientific prestige}.
\newblock
\newblock
\showeprint[arxiv]{0912.4141}~[cs.DL]


\bibitem[\protect\citeauthoryear{He, Zhang, Ren, and Sun}{He
  et~al\mbox{.}}{2015}]%
        {he2015deep}
\bibfield{author}{\bibinfo{person}{Kaiming He}, \bibinfo{person}{Xiangyu
  Zhang}, \bibinfo{person}{Shaoqing Ren}, {and} \bibinfo{person}{Jian Sun}.}
  \bibinfo{year}{2015}\natexlab{}.
\newblock \bibinfo{title}{Deep Residual Learning for Image Recognition}.
\newblock
\newblock
\showeprint[arxiv]{1512.03385}~[cs.CV]


\bibitem[\protect\citeauthoryear{Hirsch}{Hirsch}{2005}]%
        {Hirsch2005H-index}
\bibfield{author}{\bibinfo{person}{Jorge~E. Hirsch}.}
  \bibinfo{year}{2005}\natexlab{}.
\newblock \showarticletitle{An index to quantify an individual's scientific
  research output}.
\newblock \bibinfo{journal}{\emph{Proc. Natl. Acad. Sci. USA}}
  \bibinfo{volume}{102} (\bibinfo{year}{2005}), \bibinfo{pages}{16569--16572}.
\newblock


\bibitem[\protect\citeauthoryear{Hoang, Bihorac, and Rouces}{Hoang
  et~al\mbox{.}}{2019}]%
        {Hoang2019AspectBasedSA}
\bibfield{author}{\bibinfo{person}{Mickel Hoang}, \bibinfo{person}{Oskar~Alija
  Bihorac}, {and} \bibinfo{person}{Jacobo Rouces}.}
  \bibinfo{year}{2019}\natexlab{}.
\newblock \showarticletitle{Aspect-Based Sentiment Analysis using BERT}. In
  \bibinfo{booktitle}{\emph{NODALIDA}}.
\newblock


\bibitem[\protect\citeauthoryear{Hu, Peng, Huang, Li, and Lv}{Hu
  et~al\mbox{.}}{2019}]%
        {hu-etal-2019-open}
\bibfield{author}{\bibinfo{person}{Minghao Hu}, \bibinfo{person}{Yuxing Peng},
  \bibinfo{person}{Zhen Huang}, \bibinfo{person}{Dongsheng Li}, {and}
  \bibinfo{person}{Yiwei Lv}.} \bibinfo{year}{2019}\natexlab{}.
\newblock \showarticletitle{Open-Domain Targeted Sentiment Analysis via
  Span-Based Extraction and Classification}. In
  \bibinfo{booktitle}{\emph{Proceedings of the 57th Annual Meeting of the
  Association for Computational Linguistics}}. \bibinfo{publisher}{Association
  for Computational Linguistics}, \bibinfo{address}{Florence, Italy},
  \bibinfo{pages}{537--546}.
\newblock
\urldef\tempurl%
\url{https://doi.org/10.18653/v1/P19-1051}
\showDOI{\tempurl}


\bibitem[\protect\citeauthoryear{Jurgens, Kumar, Hoover, McFarland, and
  Jurafsky}{Jurgens et~al\mbox{.}}{2016}]%
        {Jurgens2016CitationCF}
\bibfield{author}{\bibinfo{person}{David Jurgens}, \bibinfo{person}{Srijan
  Kumar}, \bibinfo{person}{Raine Hoover}, \bibinfo{person}{Daniel~A.
  McFarland}, {and} \bibinfo{person}{Dan Jurafsky}.}
  \bibinfo{year}{2016}\natexlab{}.
\newblock \showarticletitle{Citation Classification for Behavioral Analysis of
  a Scientific Field}.
\newblock \bibinfo{journal}{\emph{ArXiv}}  \bibinfo{volume}{abs/1609.00435}
  (\bibinfo{year}{2016}).
\newblock


\bibitem[\protect\citeauthoryear{Kaplan, Iida, and Tokunaga}{Kaplan
  et~al\mbox{.}}{2009}]%
        {kaplan2009automatic}
\bibfield{author}{\bibinfo{person}{Dain Kaplan}, \bibinfo{person}{Ryu Iida},
  {and} \bibinfo{person}{Takenobu Tokunaga}.} \bibinfo{year}{2009}\natexlab{}.
\newblock \showarticletitle{Automatic extraction of citation contexts for
  research paper summarization: A coreference-chain based approach}. In
  \bibinfo{booktitle}{\emph{Proceedings of the 2009 Workshop on Text and
  Citation Analysis for Scholarly Digital Libraries (NLPIR4DL)}}.
  \bibinfo{pages}{88--95}.
\newblock


\bibitem[\protect\citeauthoryear{Kaplan, Tokunaga, and Teufel}{Kaplan
  et~al\mbox{.}}{2016}]%
        {DainKaplan2016}
\bibfield{author}{\bibinfo{person}{Dain Kaplan}, \bibinfo{person}{Takenobu
  Tokunaga}, {and} \bibinfo{person}{Simone Teufel}.}
  \bibinfo{year}{2016}\natexlab{}.
\newblock \showarticletitle{Citation Block Determination Using Textual
  Coherence}.
\newblock \bibinfo{journal}{\emph{Journal of Information Processing}}
  \bibinfo{volume}{24}, \bibinfo{number}{3} (\bibinfo{year}{2016}),
  \bibinfo{pages}{540--553}.
\newblock
\urldef\tempurl%
\url{https://doi.org/10.2197/ipsjjip.24.540}
\showDOI{\tempurl}


\bibitem[\protect\citeauthoryear{Liu}{Liu}{2017}]%
        {Liu2017SentimentAO}
\bibfield{author}{\bibinfo{person}{Haixia Liu}.}
  \bibinfo{year}{2017}\natexlab{}.
\newblock \showarticletitle{Sentiment Analysis of Citations Using Word2vec}.
\newblock \bibinfo{journal}{\emph{ArXiv}}  \bibinfo{volume}{abs/1704.00177}
  (\bibinfo{year}{2017}).
\newblock


\bibitem[\protect\citeauthoryear{Luo, Ji, Li, Jiang, and Duan}{Luo
  et~al\mbox{.}}{2020}]%
        {luo-etal-2020-grace}
\bibfield{author}{\bibinfo{person}{Huaishao Luo}, \bibinfo{person}{Lei Ji},
  \bibinfo{person}{Tianrui Li}, \bibinfo{person}{Daxin Jiang}, {and}
  \bibinfo{person}{Nan Duan}.} \bibinfo{year}{2020}\natexlab{}.
\newblock \showarticletitle{{GRACE}: Gradient Harmonized and Cascaded Labeling
  for Aspect-based Sentiment Analysis}. In \bibinfo{booktitle}{\emph{Findings
  of the Association for Computational Linguistics: EMNLP 2020}}.
  \bibinfo{publisher}{Association for Computational Linguistics},
  \bibinfo{address}{Online}, \bibinfo{pages}{54--64}.
\newblock
\urldef\tempurl%
\url{https://doi.org/10.18653/v1/2020.findings-emnlp.6}
\showDOI{\tempurl}


\bibitem[\protect\citeauthoryear{Ma, Xu, and Zhang}{Ma et~al\mbox{.}}{2018}]%
        {Ma-2018-AICTS}
\bibfield{author}{\bibinfo{person}{Shutian Ma}, \bibinfo{person}{Jin Xu}, {and}
  \bibinfo{person}{Chengzhi Zhang}.} \bibinfo{year}{2018}\natexlab{}.
\newblock \showarticletitle{Automatic Identification of Cited Text Spans: A
  Multi-Classifier Approach over Imbalanced Dataset}.
\newblock \bibinfo{journal}{\emph{Scientometrics}} \bibinfo{volume}{116},
  \bibinfo{number}{2} (\bibinfo{date}{aug} \bibinfo{year}{2018}),
  \bibinfo{pages}{1303–1330}.
\newblock
\showISSN{0138-9130}
\urldef\tempurl%
\url{https://doi.org/10.1007/s11192-018-2754-2}
\showDOI{\tempurl}


\bibitem[\protect\citeauthoryear{Milstead}{Milstead}{1980}]%
        {Milstead1980CitationIT}
\bibfield{author}{\bibinfo{person}{Jessica~L. Milstead}.}
  \bibinfo{year}{1980}\natexlab{}.
\newblock \showarticletitle{Citation Indexing—Its Theory and Application in
  Science, Technology and Humanities. Wiley, Oxford (1979), 274, \$15.95}.
\newblock \bibinfo{journal}{\emph{Information Processing and Management}}
  \bibinfo{volume}{16} (\bibinfo{year}{1980}).
\newblock


\bibitem[\protect\citeauthoryear{Moed}{Moed}{2010}]%
        {MOED2010265}
\bibfield{author}{\bibinfo{person}{Henk~F. Moed}.}
  \bibinfo{year}{2010}\natexlab{}.
\newblock \showarticletitle{Measuring contextual citation impact of scientific
  journals}.
\newblock \bibinfo{journal}{\emph{Journal of Informetrics}}
  \bibinfo{volume}{4}, \bibinfo{number}{3} (\bibinfo{year}{2010}),
  \bibinfo{pages}{265--277}.
\newblock
\showISSN{1751-1577}
\urldef\tempurl%
\url{https://doi.org/10.1016/j.joi.2010.01.002}
\showDOI{\tempurl}


\bibitem[\protect\citeauthoryear{Page, Brin, Motwani, and Winograd}{Page
  et~al\mbox{.}}{1999}]%
        {Page-1999-Pagerank}
\bibfield{author}{\bibinfo{person}{Lawrence Page}, \bibinfo{person}{Sergey
  Brin}, \bibinfo{person}{Rajeev Motwani}, {and} \bibinfo{person}{Terry
  Winograd}.} \bibinfo{year}{1999}\natexlab{}.
\newblock \bibinfo{booktitle}{\emph{The PageRank Citation Ranking: Bringing
  Order to the Web.}}
\newblock \bibinfo{type}{Technical Report} 1999-66.
  \bibinfo{institution}{Stanford InfoLab}.
\newblock
\urldef\tempurl%
\url{http://ilpubs.stanford.edu:8090/422/}
\showURL{%
\tempurl}
\newblock
\shownote{Previous number = SIDL-WP-1999-0120}.


\bibitem[\protect\citeauthoryear{Qazvinian and Radev}{Qazvinian and
  Radev}{2010}]%
        {qazvinian2010identifying}
\bibfield{author}{\bibinfo{person}{Vahed Qazvinian} {and}
  \bibinfo{person}{Dragomir Radev}.} \bibinfo{year}{2010}\natexlab{}.
\newblock \showarticletitle{Identifying Non-Explicit Citing Sentences for
  Citation-Based Summarization.}. In \bibinfo{booktitle}{\emph{Proceedings of
  the 48th annual meeting of the association for computational linguistics}}.
  \bibinfo{pages}{555--564}.
\newblock


\bibitem[\protect\citeauthoryear{Ruder, Ghaffari, and Breslin}{Ruder
  et~al\mbox{.}}{2016}]%
        {Ruder2016AHM}
\bibfield{author}{\bibinfo{person}{Sebastian Ruder}, \bibinfo{person}{Parsa
  Ghaffari}, {and} \bibinfo{person}{John~G. Breslin}.}
  \bibinfo{year}{2016}\natexlab{}.
\newblock \showarticletitle{A Hierarchical Model of Reviews for Aspect-based
  Sentiment Analysis}. In \bibinfo{booktitle}{\emph{EMNLP}}.
\newblock


\bibitem[\protect\citeauthoryear{Seglen}{Seglen}{1997}]%
        {Seglen1997WhyTI}
\bibfield{author}{\bibinfo{person}{Per~O. Seglen}.}
  \bibinfo{year}{1997}\natexlab{}.
\newblock \showarticletitle{Why the impact factor of journals should not be
  used for evaluating research}.
\newblock \bibinfo{journal}{\emph{BMJ}}  \bibinfo{volume}{314}
  (\bibinfo{year}{1997}), \bibinfo{pages}{497}.
\newblock


\bibitem[\protect\citeauthoryear{Sun, Huang, and Qiu}{Sun
  et~al\mbox{.}}{2019}]%
        {Sun2019UtilizingBF}
\bibfield{author}{\bibinfo{person}{Chi Sun}, \bibinfo{person}{Luyao Huang},
  {and} \bibinfo{person}{Xipeng Qiu}.} \bibinfo{year}{2019}\natexlab{}.
\newblock \showarticletitle{Utilizing BERT for Aspect-Based Sentiment Analysis
  via Constructing Auxiliary Sentence}. In \bibinfo{booktitle}{\emph{NAACL}}.
\newblock


\bibitem[\protect\citeauthoryear{Teufel, Siddharthan, and Tidhar}{Teufel
  et~al\mbox{.}}{2006}]%
        {Teufel2006AutomaticCO}
\bibfield{author}{\bibinfo{person}{Simone Teufel}, \bibinfo{person}{Advaith
  Siddharthan}, {and} \bibinfo{person}{Dan Tidhar}.}
  \bibinfo{year}{2006}\natexlab{}.
\newblock \showarticletitle{Automatic classification of citation function}. In
  \bibinfo{booktitle}{\emph{EMNLP}}.
\newblock


\bibitem[\protect\citeauthoryear{Trusca, Wassenberg, Frasincar, and
  Dekker}{Trusca et~al\mbox{.}}{2020}]%
        {Trusca2020AHA}
\bibfield{author}{\bibinfo{person}{Maria~Mihaela Trusca}, \bibinfo{person}{Daan
  Wassenberg}, \bibinfo{person}{Flavius Frasincar}, {and}
  \bibinfo{person}{Rommert Dekker}.} \bibinfo{year}{2020}\natexlab{}.
\newblock \showarticletitle{A Hybrid Approach for Aspect-Based Sentiment
  Analysis Using Deep Contextual Word Embeddings and Hierarchical Attention}.
  In \bibinfo{booktitle}{\emph{ICWE}}.
\newblock


\bibitem[\protect\citeauthoryear{Valenzuela, Ha, and Etzioni}{Valenzuela
  et~al\mbox{.}}{2015}]%
        {Valenzuela2015IdentifyingMC}
\bibfield{author}{\bibinfo{person}{Marco Valenzuela}, \bibinfo{person}{Vu~A.
  Ha}, {and} \bibinfo{person}{Oren Etzioni}.} \bibinfo{year}{2015}\natexlab{}.
\newblock \showarticletitle{Identifying Meaningful Citations}. In
  \bibinfo{booktitle}{\emph{AAAI Workshop: Scholarly Big Data}}.
\newblock


\bibitem[\protect\citeauthoryear{Wallaart and Frasincar}{Wallaart and
  Frasincar}{2019}]%
        {Wallaart2019AHA}
\bibfield{author}{\bibinfo{person}{Olaf Wallaart} {and}
  \bibinfo{person}{Flavius Frasincar}.} \bibinfo{year}{2019}\natexlab{}.
\newblock \showarticletitle{A Hybrid Approach for Aspect-Based Sentiment
  Analysis Using a Lexicalized Domain Ontology and Attentional Neural Models}.
  In \bibinfo{booktitle}{\emph{ESWC}}.
\newblock


\bibitem[\protect\citeauthoryear{Wang, Lu, and Zhai}{Wang
  et~al\mbox{.}}{2010}]%
        {Wang2010LatentAR}
\bibfield{author}{\bibinfo{person}{Hongning Wang}, \bibinfo{person}{Yue Lu},
  {and} \bibinfo{person}{ChengXiang Zhai}.} \bibinfo{year}{2010}\natexlab{}.
\newblock \showarticletitle{Latent aspect rating analysis on review text data:
  a rating regression approach}.
\newblock \bibinfo{journal}{\emph{Proceedings of the 16th ACM SIGKDD
  international conference on Knowledge discovery and data mining}}
  (\bibinfo{year}{2010}).
\newblock


\bibitem[\protect\citeauthoryear{Xu, Liu, Shu, and Yu}{Xu
  et~al\mbox{.}}{2019}]%
        {Xu2019BERTPF}
\bibfield{author}{\bibinfo{person}{Hu Xu}, \bibinfo{person}{Bing Liu},
  \bibinfo{person}{Lei Shu}, {and} \bibinfo{person}{Philip~S. Yu}.}
  \bibinfo{year}{2019}\natexlab{}.
\newblock \showarticletitle{BERT Post-Training for Review Reading Comprehension
  and Aspect-based Sentiment Analysis}. In \bibinfo{booktitle}{\emph{NAACL}}.
\newblock


\bibitem[\protect\citeauthoryear{Zerva, quoc Nghiem, Nguyen, and
  Ananiadou}{Zerva et~al\mbox{.}}{2020}]%
        {Chrysoula-2020-CTSI}
\bibfield{author}{\bibinfo{person}{Chrysoula Zerva}, \bibinfo{person}{Minh quoc
  Nghiem}, \bibinfo{person}{{Nhung T. H.} Nguyen}, {and}
  \bibinfo{person}{Sophia Ananiadou}.} \bibinfo{year}{2020}\natexlab{}.
\newblock \showarticletitle{Cited text span identification for scientific
  summarisation using pre-trained encoders}.
\newblock \bibinfo{journal}{\emph{Scientometrics}} (\bibinfo{date}{7 May}
  \bibinfo{year}{2020}).
\newblock
\showISSN{0138-9130}
\urldef\tempurl%
\url{https://doi.org/10.1007/s11192-020-03455-z}
\showDOI{\tempurl}


\end{thebibliography}
\end{document}